# Recent Developments at the Boquete Optical SETI Observatory and Owl Observatory

Marlin N. Schuetz (Boquete Optical SETI Observatory, Boquete, Panama )

**Abstract**

Progress at the privately owned Boquete Optical SETI Observatory in Panama and the Owl Observatory in Michigan is reported. The Boquete Observatory has been dedicated to the development of innovative optical SETI detectors and observations since 2010. It is currently equipped with a 0.5 meter Newtonian main telescope and a piggybacked 0.35 meter Cassegrain for tracking. Although small, the observatory's telescope and detector system has capabilities that are equivalent to most other institutional optical SETI facilities (Schuetz, M. et al., 2016). The optical SETI detectors at Boquete have evolved through many stages from a three photomultiplier coincidence detector to the current single photomultiplier version capable of detecting pulse widths up to 50 ns and for coincidence detection against a wide range of stellar background counts. The Owl Observatory photometer has similarly been in continuous development to improve performance. These activities have sought to provide practical solutions to the needs of optical SETI.

Over the past 5 years the search for laser like signals at Boquete has included over 5000 stellar objects. Yet, until early 2017 year there were large gaps in the search parameters that limited the thoroughness of those searches. Reported herein are developments that have filled many of those gaps further extending the search boundaries. With the new capabilities, the search through the list of stellar candidates out to 200 ly was begun anew.

**Keywords:** SETI, optical SETI, instrumentation, photometer

## 1. Introduction

Since the turn of the century and the beginning of serious optical searches for laser signals, significant uncertainties have continued to exist regarding the various signal parameters that might be expected. Compared with radio SETI, it is thought that the optical regime has far fewer and more narrowly defined variables. At the present time there are four general types of signals for which detection methods have been developed: continuous wave, pulsed signal photon coincidence, the detection of pulse widths from ~1 ns to >50 ns (including coincidence detection) and most recently high repetition rate, comparatively low energy pulsed signals. A brief description of each will be helpful. The discussions are limited to the timely practical application of mature technologies.

**Continuous wave (cw) signaling** has been considered and projects examining archived stellar spectral data have been done (Marcey et al. 2002, 2017). Those projects, sensitive to megawatt class laser signals, examined more than 6000 stars for laser emissions; none were detected, but as with all other limited searches few conclusions may be drawn.

When compared with a low repetition rate pulsed laser beacon, cw signaling requires a much less powerful laser, however, at the receiving end, the telescope size, spectrum analysis and data reduction requirements are more elaborate. To mitigate this requirement archival data mining has been favored. Unlike other methods for signal detection, search techniques for narrow bandwidth signals can achieve high sensitivity using large aperture telescopes. CW transmission is not conducive to the simultaneous targeting of a large number of stellar systems.

**Coincidence photon detection** or pulse pileup has been the approach favored by most institutions for the past two decades. Its first application is believed to have been at Harvard University (Horowitz et al., 2001). In this technique, the incoming optical beam is split into two or more paths with each path having a separate photodetector. If a pulse is simultaneously detected in two or more detectors a coincident event is registered. This method provides a good means of differentiating a signal from high stellar background count rates. Yet, because of the optical beam division and other losses, the detector system has significantly lower sensitivity than the single photodetector method yet to be described. If periodic pulses are the expected means of gaining attention, then observations with less than optimum sensitivity seems counterproductive. The multichannel coincidence detector is somewhat complicated, expensive and ignores signals having pulse lengths greater than a few nanoseconds in length.

**High pulse repetition rate, low power signaling** as suggested by Leeb, W. et al., 2013, is a recent addition to interstellar laser beacon possibilities. In this scheme comparatively low power lasers may transmit pulses at high repetition rates, i.e., a hundred Hertz to megahertz, but with low duty cycles. For the reception of such signals it was shown that by time stamping all photodetector output pulses and determining each pulse's time relationship with all other pulses in an interval it was possible to detect single photon periodic events against large background counts. If the technique can be shown to have both transmission and detection advantages, it may broaden an otherwise relatively restricted set of parameters for optical SETI. The method may be difficult to implement for targeting large numbers of exoplanets.

**Longer pulse length, low repetition rate signal detection** is the approach that has been pursued at Boquete for the past 6 years and will soon be employed at the Owl Observatory. In this paper we consider that a pulsed laser beacon, having pulse lengths ranging from ~1 to 50 ns, may be favored over other techniques. The method is discussed in detail and includes transmission characteristics and advantages, the various detection parameters, two detector versions (analog and digital signal discrimination), methods for determining the overall sensitivity, periodic pulse detection from within the event stream and the correlation of test results with Poisson statistics.

**Receiving telescope requirements.** One might assume that very large telescopes could be used to advantage, however, two important reasons argue otherwise.

1. Using today's technology and without optical attenuation or bandwidth limiting filters, a large telescope's photometer would become flux limited with most of the candidate stars within 500 light years. As an example, even a small 0.5 meter telescope's detector, without attenuating or bandwidth filters, can become flux limited with stars brighter than about 6$^{th}$ magnitude. That same telescope can observe and collect meaningful real time data on stars to greater than 14$^{th}$ magnitude. Attenuating filters can extend a small telescope detector's useful range to include almost the entirety of local stars systems. Furthermore, telescopes somewhat smaller than 0.5 meters may have only a small sensitivity handicap in SETI searches.

2. Civilizations targeting other star systems may be concerned about the number of receiving systems a putative civilization might have. Considering stellar observational dwell times up to 20 minutes, even if not flux limited, a few large telescopes could not timely observe the hundreds of thousands of stars systems that may be necessary to achieve a successful detection. Thus, it seems reasonable that transmission facilities would provide signals capable of being detected by lesser, presumably more numerous facilities.

For this type of signal beacon, the arguments in 1 and 2 represent potentially constraining limits for the required transmission pulse energy. For example, the required laser pulse energy, $E=\Phi z^2 \lambda hc/\pi \omega_0^2$, for an 8 meter transmitting telescope aperture ($\omega_0$=4.0 m), a minimum detectable signal ($\Phi$) of 100 photons m$^{-2}$ and a maximum distance (z, meters) is approximately 0.2, 0.8 and 1.8 Megajoules for distances of 100, 200 and 300 light years respectively.

It can be shown that the radius of the illuminated disk at the target distance is $\omega_z = z \cdot \lambda / \pi \omega_0$. From that relationship, if a targeted exoplanet's position were known with sufficient precision, the transmitting system's parameters may be adjusted to minimize the beam divergence and reduce the required pulsed energy.

**The definition, justification and detection means for periodic pulsed laser signals having pulse lengths up to 50 nanoseconds.** An obvious absolute minimum detectable limit for a single laser pulse is one detected photon for any given signal collection area. That, however, falls short of practical reality wherein repetition rates, quantum efficiency, optical efficiency and stellar background levels are large factors for making such a determination.

Although several new types of devices show promise, photomultipliers (pmt) and avalanche photodiodes (apd or spad) are currently preferred for astronomical single photon detection. Common side-on photomultipliers have peak quantum efficiencies (QE) up to 35 percent and typically about 190 mm$^2$ effective photocathode area. Avalanche photodiodes can have peak quantum efficiencies from 40 to 80+ percent, but have much smaller areas for light detection. For various practical reasons the Hamamatsu R3896 photomultiplier was chosen for use at the Boquete and Owl facilities. This device has a spectral response range of 185 to 900 nm with a peak quantum efficiency of 30%. For these discussions, however, a quantum efficiency of 20 percent has been chosen as a nominal figure; thus, approximately one out of five photons impinging on the pmt's photocathode will be consistently detected.

**The detection of periodic signals.** We next examine the process by which laser signals having a low repetition rate (<<1 cps) may be detected against large stellar backgrounds, e.g. up to ~5x10$^5$ counts per second. Here, the detector design takes advantage of Poisson statistics wherein the Poisson rate is: $R = r^n \cdot t^{(n-1)}/(n-1)!$. With a random pulse arrival rate (r), the equation describes the rate of "hits" (R) having n events occurring within the interval t (for r·t<<1) (Meade, C.C., 2013).

The photometer detector has several discriminators. A first discriminator excludes the photomultiplier low level noise pulses. A second discriminator, either analog or digital, discards all single photoelectron pulses leaving only the selected Poisson rate noise pulses and other signals. Yet a third (coincidence) discriminator samples the pmt output and discards all pulses that are not at least twice the amplitude of a typical single pulse event.

For our SETI photometers the interval (t) is small, i.e., 1 to 50 ns, and nearly all of the random stellar photoelectron pulses may be eliminated leaving only those unusual events having 2 or more detected photons within the interval. Photons bunched in time from laser pulses fall into this category. The second discriminator threshold is automatically adjusted according to the stellar background count rate.

From the previous example's requirement of 5 photons as a minimum for a consistent single photon detection within the optical collection area, we now require that at least another 5 photons are present to produce one additional detected photon within the time interval, i.e., for n=2. And finally, taking a typical telescope's ~75% optical efficiency the minimum detectable limit is elevated to about 13 photons. For the 0.5 meter Boquete telescope, that minimum becomes about 65 photons m$^{-2}$ at the telescope's aperture.

With some optimism, it may be reasonable to expect that our ET friends would err on the conservative side and provide more energy per pulse. However, since the energy requirement for these transmissions is substantial, it seems unlikely they would exceed a few hundred photons m$^{-2}$ by an unnecessarily large factor. Using our economic standards and a 20% mains efficiency the energy cost for each one megajoule pulse would be about $0.25. If an earth-based system transmitted 1 pulse sec$^{-1}$ the annual cost for energy would be about $8 million. A transmitting satellite may be expected to be energy independent and likely more constrained by overall project costs, targeting time, cooling and laser damage issues.

**Optical Band Considerations.** At this time in our optical SETI efforts, automated space based detection systems are only a distant dream. Thus, for the present we limit the discussion to optical wavelengths that can be detected at ground level and include the detector's entire spectral response range.

For small receiving telescopes, atmospheric extinction mostly rules out searching at wavelengths shorter than about 350 nm; above 450 nm the atmospheric losses are relatively small. For distances less than about 1000 ly, laser pulses in the optical spectrum would pass through the interstellar medium (ISM) mostly unimpeded. Beyond 1000 ly, near-infrared (NIR) would be favored since it is little attenuated by the interstellar medium (ISM). One might, therefore, consider opting for NIR detectors to allow searches at all distances. Note, however, that as the transmitting laser's wavelength is increased, the aperture radius of a diffraction limited transmitting telescope must increase proportionally to compensate for beam's growing whole angle of divergence. That is, $\theta = \lambda/\pi\omega_0$, where $\omega_0$ is the beam waist at its narrowest point, i.e. the transmitting telescope's aperture. To maintain the same receiving end total signal flux, the requirement for an 8 meter telescope at 500 nm would become a 16 meter telescope for transmission at 1 um. Detector noise is also necessarily greater at longer wavelengths.

**Laser transmitter parameter limits**. Roughly known are the energy and pulse rates needed for an efficacious low repetition rate beacon. These values aid in speculating about the limits of the other parameters. For instance, each optical device within the beam must survive the combination of laser transmission wavelength, fluence, pulse repetition rate and thermal extremes. Although the damage threshold for mirrors and lenses may be mitigated with advanced materials there must exist limits for that which is reasonable. A commonly used rule of thumb for approximating the laser induced damage threshold (LIDT) for pulse widths in the low nanosecond range is:

$$LIDT_2 \sim LIDT_1(\lambda_2/\lambda_1)\sqrt{(t_2/t_1)}$$

Where $t_1$ is the laser pulse length and $\lambda_1$ is the laser wavelength for a known $LIDT_1$, and $t_2$ is the laser pulse length and $\lambda_2$ is the laser wavelength for the unknown $LIDT_2$. The approximation is presented only as a indicator of the relationship of wavelength and pulse width as regards the damage threshold. It suggests that for a given pulse fluence, longer wavelengths and pulse widths are favored to elevate the threshold for damage. The larger optical areas required with longer wavelengths also aid in that regard. It is reasonable to expect that advanced materials and techniques can elevate the LIDT beyond that which is possible given our current understanding (Zaghloul, 2002). The difficulties at each extreme of the optical SETI useful spectrum suggest there may be a convenient middle ground from 500-1000nm.

Often discussed is the use of narrow bandwidth detectors to improve the signal to background relationship. For established interstellar communications that technique can be used to great advantage. For SETI, however, without large expenditures for equipment and observing time a low energy narrow band pulsed beacon may be easily overlooked.

**Pulse Length.** Nearly all of the previous optical SETI detection methods have been designed for multiple photodetectors in what has become known as the coincident detection method. Although individual detectors may easily detect large amplitude (coincident) pulses, the technique has not been favored since they also react to a variety of phenomena including gamma radiation and corona micro-discharges. Those phenomena can produce large amplitude detector output pulses and result in false positive detections.

Less desirable features are that the coincident detection scheme: ignores the possibility of longer pulse lengths, has reduced sensitivity due to losses in the additional optical elements and the perceived need to minimize false positives via high detection threshold levels (reduced sensitivity).

A second method having just a single photodetector is current used at the Boquete observatory and will soon be installed at the Owl Observatory. The photometer triggers an output when either a large amplitude (coincident-like) signal occurs and/or when two or more photon detections occur within about 50 ns. While the arguments against the single photodetector method would be valid for non-periodic signal detection, those issues are entirely mitigated for periodic signals through the application of downstream hardware and software. The particular benefits of this method are the ability to detect both periodic coincident and non-coincident events at high sensitivity along with simplicity and reduced cost. Further advantages of longer length pulses are: reduced laser peak power requirements and a signal may serve as both a beacon and to convey coded data.

Advantageously, the pulse length range also has self limiting features. That is, as the pulse length is reduced, greater laser peak power is required to achieve the energy needed per pulse, e.g., a pulse 1 ns in length having 2 megajoules energy requires 2 petawatts peak power, but at 50 ns the peak power is reduced to 40 gigawatts. At the other extreme, pulse lengths are limited by a reduction in the signal to background count ratio, i.e., as the detection interval is lengthened more stellar photons slip through the discriminator as predicted by Poisson statistics. Thus, ~100 ns may be a current practical signal detection interval limit. Note however, that were longer pulse lengths to be encountered, it is possible that sufficient photons would fall within the first 50 ns to trigger a detection. If sufficiently energetic, pulses shorter than 1 ns would also be detectable.

As mentioned, long pulse lengths offer the unique opportunity to convey a data stream. For example, a long pulse may be a composite of sequentially varied wavelengths or of multiple short pulses using any of various coding techniques, e.g. pulse position modulation (PPM) or pulse width modulation (PWM), without degrading the ability to detect the beacon signal as a single long pulse.

At the Boquete and Owl Observatories system performance is routinely tested using a pulsed LED method that provides proof of detection for single photons, coincident photons and longer pulse lengths (to 50 ns) having 2 or more detected photons. Advantageously the pulsed LED has precise, crystal controlled, periodicities ranging from 2 to 0.002 Hz. A second dc powered LED provides a simulated stellar background with detected photon emission rates from 0-1 Mcps. Thus, the detection sensitivity for periodic pulsed signals can be tested against large background counts. During all stellar observations, a low level 64 second periodic test pulse runs continuously as proof of performance.

A further note about test signals will be useful. At Boquete, the two LED light signals are coupled into a 2 mm poly(methyl methacrylate) PMMA flexible side glow fiber via two tight radius bends. A length of the fiber is located at the photomultiplier in an otherwise unused area of the glass envelope adjacent to the photometer aperture and photocathode. The configuration has several advantages; the LED pulsed signal leads can be kept short for pulse purity, the area of light emitted from the fiber can be masked appropriately and the signals for both the background and pulsed signals impinge upon the same area of the photocathode. This method of introducing tests signals has been found to be more easily replicated and performs better than previous arrangements.

**Pulse Periodicity.** Considering that periodically pulsed signals provide for an easily detectable beacon, it is difficult to imagine that a signaling civilization would do otherwise.

Over the years there have been many ideas regarding the timing of these signals ranging from millions of pulses per second to less than one pulse per thousand seconds. When the various transmission and reception issues are taken into account, there are reasons to expect that the extremes for the range for high energy pulsed signals can be appreciably narrowed.

High rate pulsed signals, i.e., >1 pulse sec$^{-1}$, are clearly appealing for their ease of detection, but the need for illuminating multiple targets may be too demanding. Targeting time, energy requirements and optical component damage are the primary limiting factors for a laser transmission system. At the other extreme, although one pulse every 1000 seconds or more simplifies the transmitter design, offsetting that is the inability to target a large number of stars. Signal detection at the receiving end is also limited by the requirement that three and more pulse detections are required to confirm periodicity. While one pulse per thousand seconds is not wholly unreasonable, i.e., 100 telescopes observing 4 stars per night could cover 200,000 stars in 500 observing nights; that level of dedication might only be expected after the historical first confirmed signal detection; not dissimilar to the current exoplanet frenzy for discovery.

From the perspective of a laser system designer, there are other limiting characteristics. First, it seems safe to assume that laser transmitters would be space based, i.e., ground based systems would have atmospheric losses, pointing and safety issues. An in-depth study is needed to develop parameters that are good compromises for both the transmission and reception requirements.

**Other issues of concern.** Satellite targeting complexities bring to light yet other strategic search questions. If a single satellite can only cover a small area of the sky, how might this be incremented to achieve a broader beacon pattern? For us, that kind of long term incremental targeting translates to the requirement of repetitive searches.

Alternatively, if an alien species evolved more peaceably than homo sapiens, there might exist little need for large military budgets. To put that in perspective, if earth's collective trillion dollar per year military budget were partially available for reaching out to alien worlds many such satellites could be lofted without great fanfare. And, for a peaceful species, projects of that sort may even be needed as busy work. Clearly the range of possibilities extend far beyond these bounds.

From our perspective the uncertainty of civilizations' stability or survival over large time spans may be a strong inhibitor of attempts to initiate communications or even reply to a signal at distances greater than a few hundred light years. Just as we prioritize candidate stars it seems reasonable that any civilization invested in the transmission of signals would prioritize targets favoring their near neighbors. At earth's present level of optical SETI activities, the abundance of stars system within a few hundred light years includes a sufficient number of candidates for many years of targeted searches.

## 2. Experimental

**At Boquete**, and prior to 2017, fast Fourier transform (fft) software was used for the detection of periodic signals. However, that particular software limited the sensitive detection range to ~0.05-10 Hz. To improve on this, custom software was developed that examines the time stamped relationship of each hit (to +/-0.5 millisecond accuracy) with all other hit time relationships within an observation dwell time. The method, a variation on that described by Leeb (2013), has resulted in the unambiguous detection of a few periodic pulses in a much larger background of random pulses. Thus, the Boquete and Owl observatories now scrutinize detected pulses for periodicities that might occur from 0.005 to >1 Hz. If future arguments are persuasive, the lower limit can be extended appropriately. Pulses detected at higher rates would be obvious and easily detected. Over the past 20 years most of the optical searches have limited the stellar dwell time to only a few minutes and the all-sky surveys to much less. The preceding arguments strongly suggest that stellar dwell times should be longer than most observatories have allowed. Taking missed pulses into account, the detection of periodicity requires more than three pulses and the observational dwell time must account for this. Boquete currently uses 720 seconds as the minimum dwell time per star allowing the detection of pulse periods out to a maximum of ~200 seconds. In the data reduction, as long as multiple pulses of a periodic pulse stream are detected, missing pulses are accounted for and scored as multiples of the fundamental period.

The new signal processing method enables the Boquete and Owl systems to achieve greater sensitivity by way of lowered discriminator threshold settings, yet without having false positive detections. Although the pulse timing error band of +/- 0.5 millisecond has been found to be adequate, there are plans to reduce this substantially in the near future. That improvement will allow the second and coincidence discriminators' thresholds to be further reduced.

**Two methods of detection using ECL devices at Boquete.** Prior to 2017 the detector board used electronic circuitry employing TTL type devices. Although these performed well there were deficiencies preventing optimal performance. For example, photomultiplier output pulses vary extensively in width and amplitude. TTL devices lacked the speed to properly process these pulse variations which lead to inaccurate results, i.e., the results did not track Poisson statistics well.

Recent development projects have resulted in detector circuit boards that employ very high speed ECL devices. The first board to be developed used an analog second discriminator to differentiate multiple pulses in a 50 ns interval from other individual random pulses. A simplified circuit may be seen in figure 1.

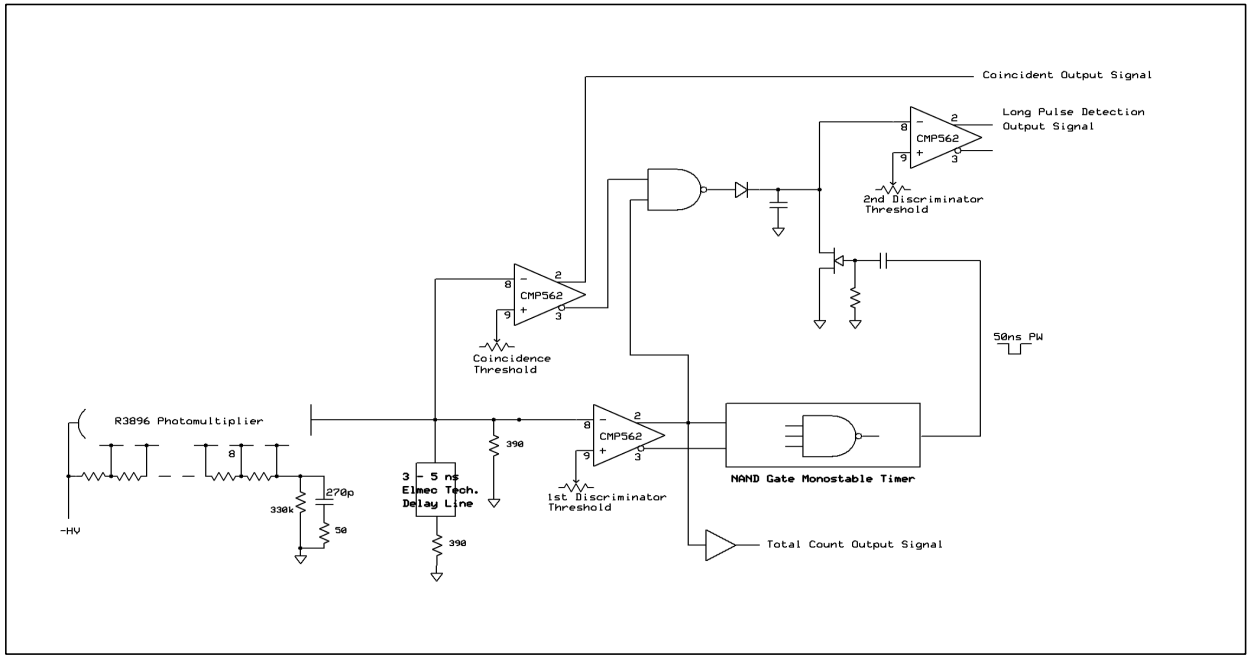

Figure 1.

There are several features of figure 1 for which explanation will be helpful. In the lower left, the photomultiplier base was modified to improve the output pulse characteristics. The capacitors that normally parallel the three lower dynode resistors were removed. A series capacitor and resistor were placed from the last dynode to ground to support high pulse rates and to help minimize pulse ringing. At the input to the first and coincidence discriminators, in lieu of a snubber, a delay line was used. Although the delay line is an inductive element and has low dc resistance an additional resistor to ground is needed to limit the pulse height. This arrangement yields well defined large amplitude pulses with little ringing. For example, the pmt, normally terminated with 50 ohms yields maximum pulse heights of about 100 mv with noise and ringing levels around 20 mv. PMT output pulse amplitudes with the above arrangement are as much a 1 volt (coincident levels) with noise and ringing levels below 50 mv. Typically single photoelectron pulses are 75 to 175 mv in amplitude. As may be seen in figure 2, the pmt delay line termination results in an exemplary pulse shape.

Figure 2

A typical pmt output pulse with 3.3 ns delay line termination (100 mv vertical and 10 ns horizonal scales).

Further improvements are expected with lesser delay line times.

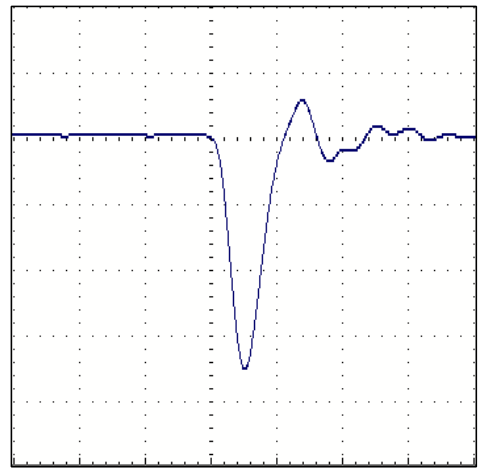

PMT output pulses are then amplitude discriminated for the two signal types, those single pulses greater than baseline noise (>75mv) and pulse pileup (coincidence) signals with large amplitudes. Highly important for proper second discriminator operation is the requirement that all first discriminator output pulses must be the same amplitude, shape and width. The output from the 1$^{st}$ discriminator very nearly achieves this requirement without additional circuitry. Remaining small variations were eliminated using a 2 ns time constant on the comparator's latching feature.

After the first discriminator, the signal branches to a monostable multivibrator that sets the Poisson interval timing and subsequently gates the integrator circuit. Each single photoelectron pulse will quickly charge the integration capacitor to a common level. If other pulses do not arrive within 50 ns, timeout occurs and the integration capacitor is discharged through the JFET. On the other hand, a second, third or fourth pulse occurring within the 50 ns interval will elevate the charge on the integration capacitor commensurate with the number of pulses. When the integration capacitor's charge level exceeds the threshold setting, the second discriminator is triggered and outputs a signal indicating a long pulse had been detected. Note that if a coincidence level pulse occurs, it is inhibited from being detected as long (group) pulses.

Because ECL signals are not compatible with downstream processes, each output signal must be converted via level translators (not shown). Downstream, TTL pulse stretchers and line drivers provide the computer interface signals.

As previously mentioned, a second generation of this circuit was developed wherein the integrator and second discriminator were replaced with a ripple counter. Multiple pulses occurring within the 50 ns interval were counted and additional gating selected the n=2, 3 or 4 operating modes. This discriminator performed near identically to the analog discriminator version, however, the inability to set continuously variable threshold levels was a predicted disadvantage. The development project usefully verified the performance of both circuits.

Figures 3a - 3f are examples of the analog discriminator operation. The top oscilloscope trace displays the integrated signal amplitude corresponding to the plurality of photoelectron pulse detections (lower trace) that occur within the 50 ns interval. The integrated signal is then tested against preset n=2, 3, 4+ threshold levels. When the threshold is exceeded, an output pulse is time stamped and the data stored. At the end of each stellar observation that data is processed exposing the 64 second period test pulse and any other periodic signals which may have occurred. It may be seen that when pulses are closely spaced, the later pulses have somewhat decreased amplitude, but there is no significant dead time following any of the pulses. The continuously variable analog discriminator threshold feature aids in achieving the highest sensitivity for laser pulsed signals balanced against the wide range of stellar background Poisson rates.

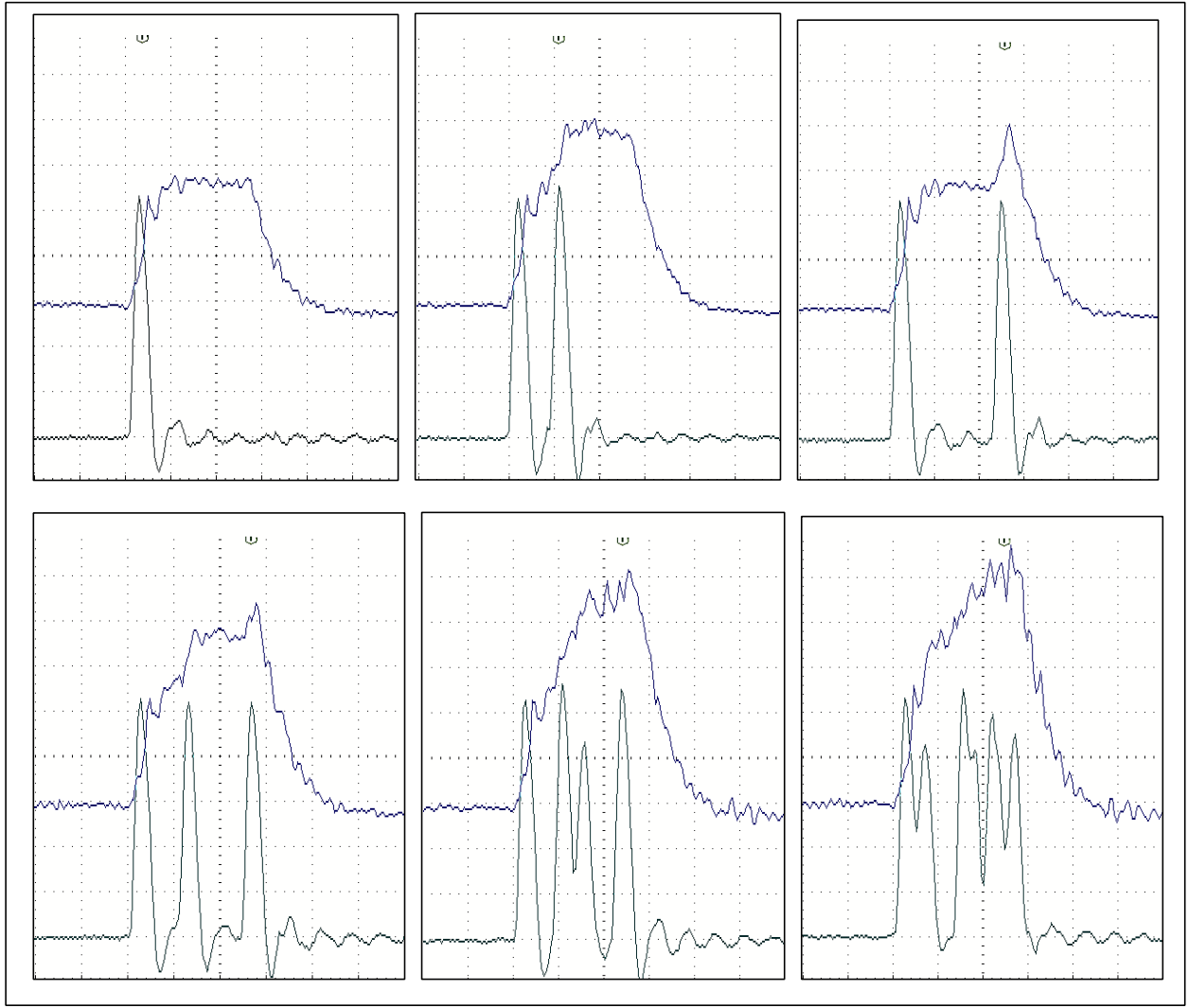

Figure 3, a,b,c,d,e,f (20ns horiz. division).

The traces in figure 3 were made using a 200 MHz bandwidth oscilloscope. As such they do not adequately display the fast rise and fall times or valleys between the peaks. A 500 MHz scope verified the salient features of figure 3, however, graphical reproductions were not possible. Such are the burdens of self funded independent research.

In Figure 4 the detected multiple pulse signals match the calculated values.  Note that at low pmt pulse rates the line representing multiple pulse detections diverges from the calculated line; in large part, cosmic ray events appear to be the cause.  Those divergent characteristics were nearly identical for both second discriminator versions.

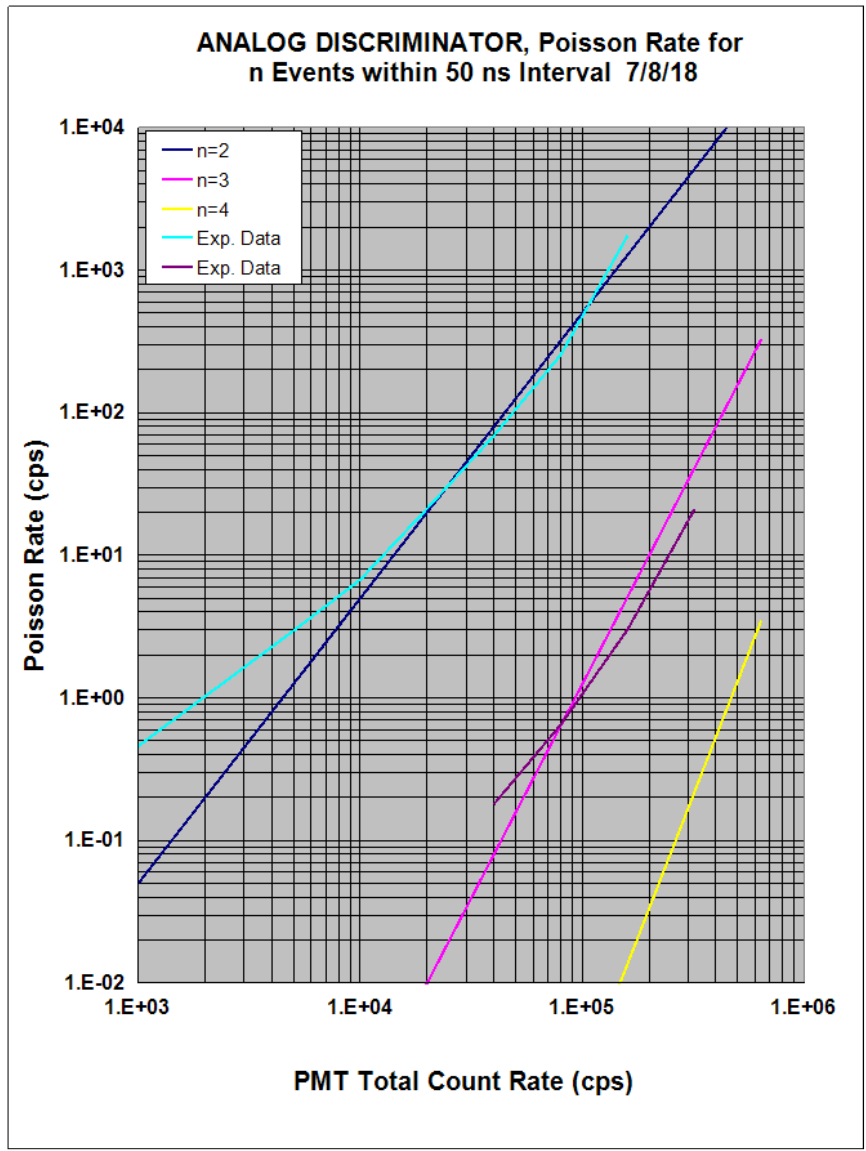

Figure 4.

The comparison of coincident event rates with Poisson statistics yields an often overlooked disparity. Note that the Poisson distribution is an appropriate model for predicting the event rates (for n=2 ,3 ,4. . .) that occur within an interval, but it is not a predictor of coincident events or coincident false positive detections. The Poisson distribution equation is a means of counting unusual events and cannot account for pulse height variations as is the case for multiple pulses merged or summed into a single pulse (n=1). The coincidence detection rate is, however, highly dependent upon the coincidence discriminator threshold potential wherein summing is the dominant feature.

Two test series were done to determine the rate of coincident events versus the photomultiplier total count rate. The first series recorded the hit rate using the analog second discriminator with a narrow (10 ns) event window. The second series was similar, but the rate of coincident detections was recorded. As may be seen in figure 5, the two tests yielded nearly identical results and confirmed the distinct difference from the Poisson rate slope.

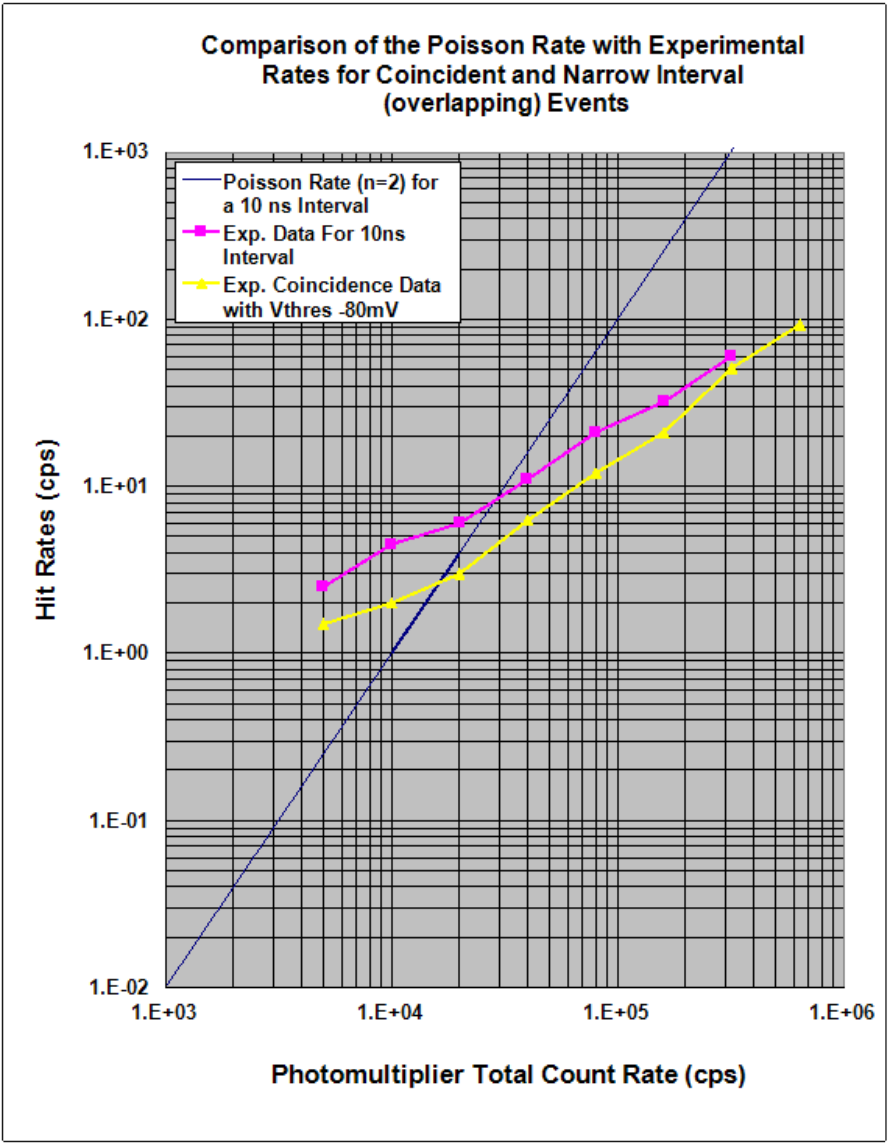

Figure 5.

Additional data similar to figure 5 revealed that, within the range of normal operation, the slopes of the data were constant, i.e., independent of the threshold voltage.

The prototype of this generation detector has been operational at Boquete since early 2018. Bruce Howard at Owl Observatory is taking the development to the next level using fabrication methods more appropriate for ultra high speed circuitry. When completed, both observatories will be so equipped and further performance improvements are expected.

**The future for the Boquete and Owl Observatories.** The Owl Observatory has been fully operational for several years. Bruce Howard will retire in August and may then dedicate full time to the observatory and its operation. The latest photometer upgrades will be installed soon as well as high accuracy time data stamping.

At the Boquete Observatory minor upgrades and system cleanup are being done during the remainder of the 2018 rainy season (until December) while opportunities for observing are few. Improvements in the timing accuracy for periodic signal detection is also a priority.

**Suggestions for other future development.** From the above discussions on signal parameters a not so daring generalization may be suggested. SETI researchers accept that there is an extremely low probability earth is timely targeted with laser signals. There is yet a smaller chance that detectors on earth will have the characteristics needed to detect such signals. Thus, to improve the odds, each class of detector should be designed to monitor the pertinent parameters as broadly as is practical; a pity, for instance, if a detectable signal has gone unnoticed because of a minor receiver deficiency. To help mitigate this issue a thorough study by appropriate experts could address the practical problems of a satellite laser transmission system. The study would be helpful in determining a probable, narrow range of signal periodicity as well as other concerns.

All-sky surveys are excellent means for rapid, thorough optical SETI searches. Shortcomings of previous surveys have been the restrictive coincidence detection method and the short dwell times inherent in non-tracking systems. In large part, these limitations can be resolved. For instance, equipping the telescope mount for limited right ascension tracking can expand the all-sky parameter space manyfold, e.g. ~10-15 minutes for each observation segment, albeit at the minor expense of extending the survey length. Detector design can employ single devices for each pixel to detect coincident and long pulses with overall improvements in sensitivity. Although detectors currently lack very broad spectral range, the front end devices could be stepwise exchanged enabling all-sky surveys to cover the optical bands of greatest interest. Additionally, detector pixels might be selectively gated for the detection only of local stars, e.g. less than 500 ly, to reduce the processing burden. With precise "hit" time stamping, the second discriminators' settings may be reduced to a minimum level and yielding high sensitivity. These are suggestions that may be accomplished at relatively small incremental costs compared with the total costs of projects with narrower search parameters. It may also be possible to retrofit existing hardware.

**Acknowledgements.** During the past 4 years, the developments at Boquete and the Owl Observatories have been collaborative. Bruce Howard's contributions to all aspects of this endeavor have been invaluable. Contact Information. We also very much appreciate the help and support of Douglas Vakoch at METI.

Marlin (Ben) Schuetz, email: mschuetz@optical-seti.org  website: http://optical-seti.org

Bruce Howard, email: bruce@owlobservatory.com  website: http://owlobservatory.com